\documentclass[prl,aps,twocolumn,showpacs]{revtex4-1}

\usepackage[centertags]{amsmath}
\usepackage{graphicx}
\usepackage{amsfonts}
\usepackage{epstopdf}
\usepackage{color}

\begin{document}

\title{Tracking breather dynamics in irregular sea state conditions}

\author{A. Chabchoub$^{1,2,\ast}$} 

\affiliation{$^1$ Department of Ocean Technology Policy and Environment, Graduate School of Frontier Sciences, The University of Tokyo, Kashiwa, Chiba 277-8563, Japan}
\affiliation{$^2$ Department of Mechanical Engineering, Aalto University, 02150 Espoo, Finland}
\email{amin.chabchoub@isea.k.u-tokyo.ac.jp}

\begin{abstract}
Breather solutions of the nonlinear Schr\"odinger equation (NLSE) are known to be considered as backbone models for extreme events in the ocean as well as in Kerr media. These exact determinisitic rogue wave (RW) prototypes on a regular background describe a wide-range of modulation instability configurations. Alternatively, oceanic or electromagnetic wave fields can be of chaotic nature and it is known that RWs may develop in such conditions as well. We report an experimental study confirming that extreme localizations in an irregular oceanic JONSWAP wave field can be tracked back to originate from exact NLSE breather solutions, such as the Peregrine breather. Numerical NLSE as well as modified NLSE simulations are both in good agreement with laboratory experiments and highlight the significance of universal weakly nonlinear evolution equations in the emergence as well as prediction of extreme events in nonlinear dispersive media. 
\end{abstract} 

\maketitle 

Ocean extreme waves, also referred to as freak or rogue waves (RWs), are known to appear without warning and having disastrous impact, in consequence of the substantial large wave heights these can reach \cite{kharif2009rogue,osborne2010nonlinear}. Studies on RWs attracted the scientific interest recently due to the interdisciplinary nature of the modulation instability (MI) of weakly nonlinear waves \cite{benjamin1967disintegration,akhmediev1985generation,zakharov2013nonlinear} as well as for the sake of accurate modeling and prediction of these mysterious extremes \cite{kharif2003physical,onorato2013roguereport,dudley2014instabilities,baronio2014vector}. Indeed, exact solutions of the nonlinear Schr\"odinger equation (NLSE) provide backbone models that can be used to describe RWs, providing therefore deterministic numerical and laboratory prototypes to reveal novel insights of MI \cite{akhmediev1997solitons}. Within the vast range of pulsating NLSE solutions on finite background, there is one prominent candidate that is known to have similar physical properties as ocean RWs, namely, the doubly-localized Peregrine breather (PB) \cite{peregrine1983water,shrira2010makes}. Despite the fact that it is theoretically assumed that the modulation period of the PB is infinite, laboratory observations confirmed that a finite number of waves in the background is sufficient to initiate its dynamics in nonlinear dispersive media \cite{kibler2010peregrine,chabchoub2011rogue,bailung2011observation}. These observations also proved that extreme localizations can be indeed discussed by means of the NLSE, despite violation of the theoretical assumption of the wave field to be or remain narrow-banded. 

Based on this latest progress, it is reasonable to study the dynamics of breathers, assuming irregularity of the underlying wave field in order to quantify limitations of the approach and to enlarge the scope of possible applications such as in oceanography. In fact, ocean waves' motion can be narrow-banded, such as in the case of swell. However, when winds, currents and wave breaking are at play, the wave field may experience strong irregularities, a state that limits applicability of the NLSE. Nevertheless, recent laboratory experiments showed the persistence of the PB in the presence of strong wind \cite{chabchoub2013experimentswind} and therefore its physical robustness to perturbations. To the best of our knowledge, the emergence of a RW in an irregular random wave field has never been tracked back to start from NLSE breather dynamics in a laboratory environment.

Here, we report an experimental study confirming the possibility for exact breather solutions to trigger extreme events in realistic oceanic conditions. According to this, the PB has been embedded into a JONSWAP field \cite{babanin2011breaking}, thus, into a realistic irregular ocean configuration with random phases in order to provide initial conditions for the experiments. In this latter hybrid surface elevation the unstable Peregrine wave packet perturbation, now cloaked in the irregular state, initiate the focusing of an extreme wave that satisfies the oceanographic definition of RW, that is, the height of the measured extreme wave indeed exceeds twice the significant wave height of the wave record. The experimental results are compared with NLSE and modified NLSE (MNLSE) predictions that show a good agreement. This certifies the possible life span of NLSE models in broad-banded processes, a fact that may be valuable in the prediction of extreme events as well as in extending the applicability range of deterministic localized structures in optics and ocean engineering. 

\begin{figure}[h]
\centering
\includegraphics[width=\columnwidth]{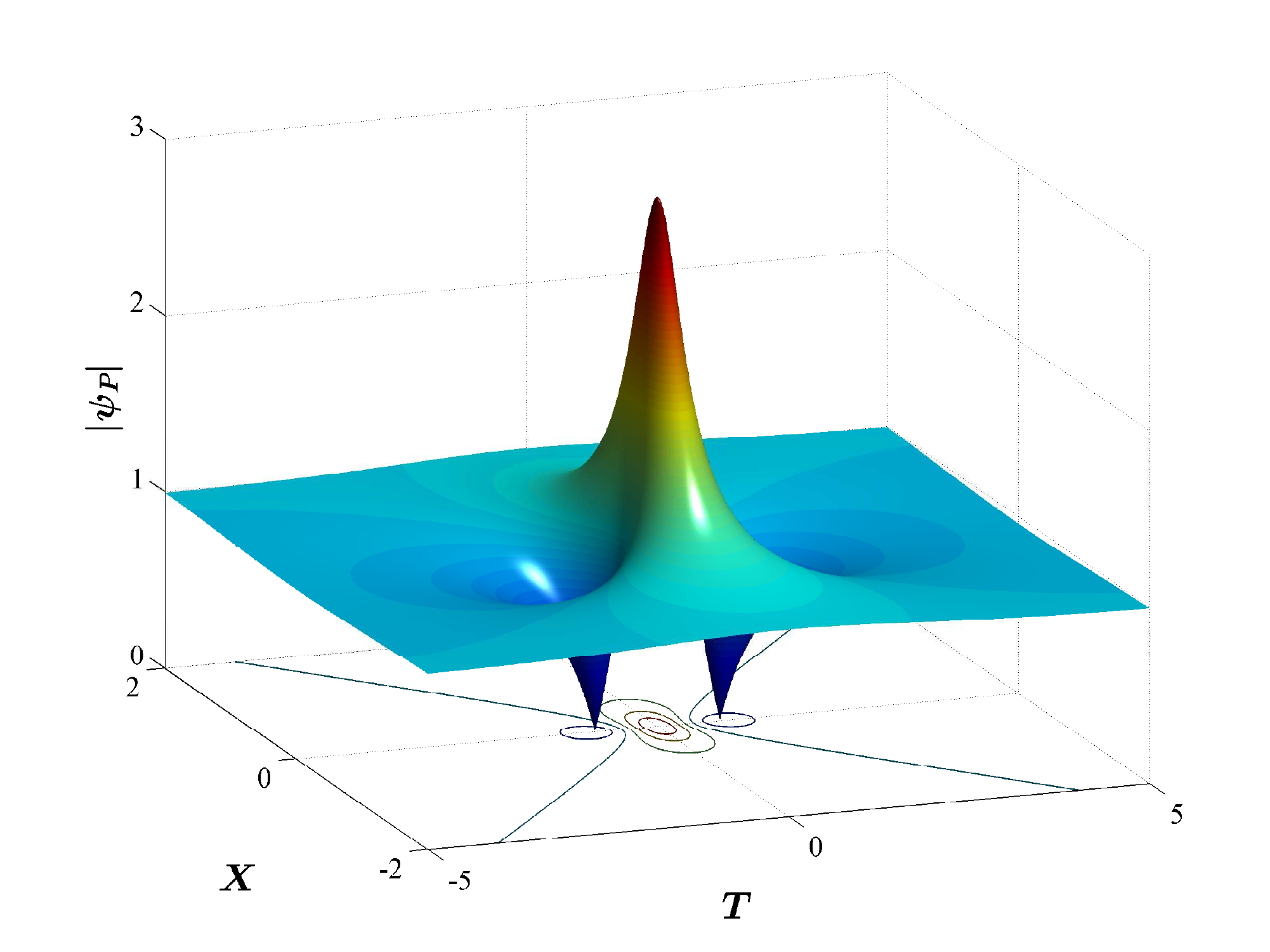}
\caption{(Color online) Doubly-localized PB solution, which enhances the amplitude regular wave field by a factor of three.}
\label{fig1}
\end{figure}

The uni-directional evolution of water wave packets $\Psi(x,t)$ can be modeled by means of the time-NLSE \cite{zakharov1968stability,osborne2010nonlinear}
\begin{equation} 
\operatorname{i}\left(\Psi_x+\dfrac{2k}{\omega}\Psi_t\right)-\dfrac{1}{g}\Psi_{tt}-k^3\left|\Psi\right|^2\Psi=0,
\label{NLS}
\end{equation}
where $g$ denotes the gravitational acceleration and the wave frequency $\omega$ is connected to the wave number $k$ through the linear dispersion relation $\omega=\sqrt{gk}$. An efficient way to model and generate a single extreme event on the water surface can be achieved by use of the PB \cite{peregrine1983water}. When considering the scaled form of the time-NLSE 
\begin{eqnarray}\textnormal{i}\psi_X+\psi_{TT}+2|\psi|^2\psi=0,
\end{eqnarray} 
this latter solution with algebraic instability growth rate reads 
\begin{eqnarray}
\psi_P(X,T)=\left(-1+\dfrac{4+16\operatorname{i}X}{1+16X^2+4T^2}\right)\exp\left(2\operatorname{i}T\right). 
\label{PB}
\end{eqnarray}
The PB solution (\ref{PB}) is depicted in Fig. \ref{fig1} while its physical properties are described in the Fig. \ref{fig1}'s caption. 
This solution is subject of intensive studies \cite{kibler2010peregrine,onorato2013rogueberlin,alberello2016velocity} due its particular physical features including the fact that it describes the MI in the case of infinite modulation period. Interestingly, this breather (or any other doubly-localized solution of this kind \cite{akhmediev1985generation,akhmediev2009rogue}) does not require an infinite number of waves in order to observe its dynamics in a physical medium \cite{chabchoub2016hydrodynamic}. Based on this fact, the aim of this study is to investigate the possibility of the PB's focusing feature to persist in chaotic conditions. To achieve this, a dimensional form of the solution is embedded in an oceanic JONSWAP wave field, as shall be described in the following. 

\begin{figure}[h]
\centering
\includegraphics[width=\columnwidth]{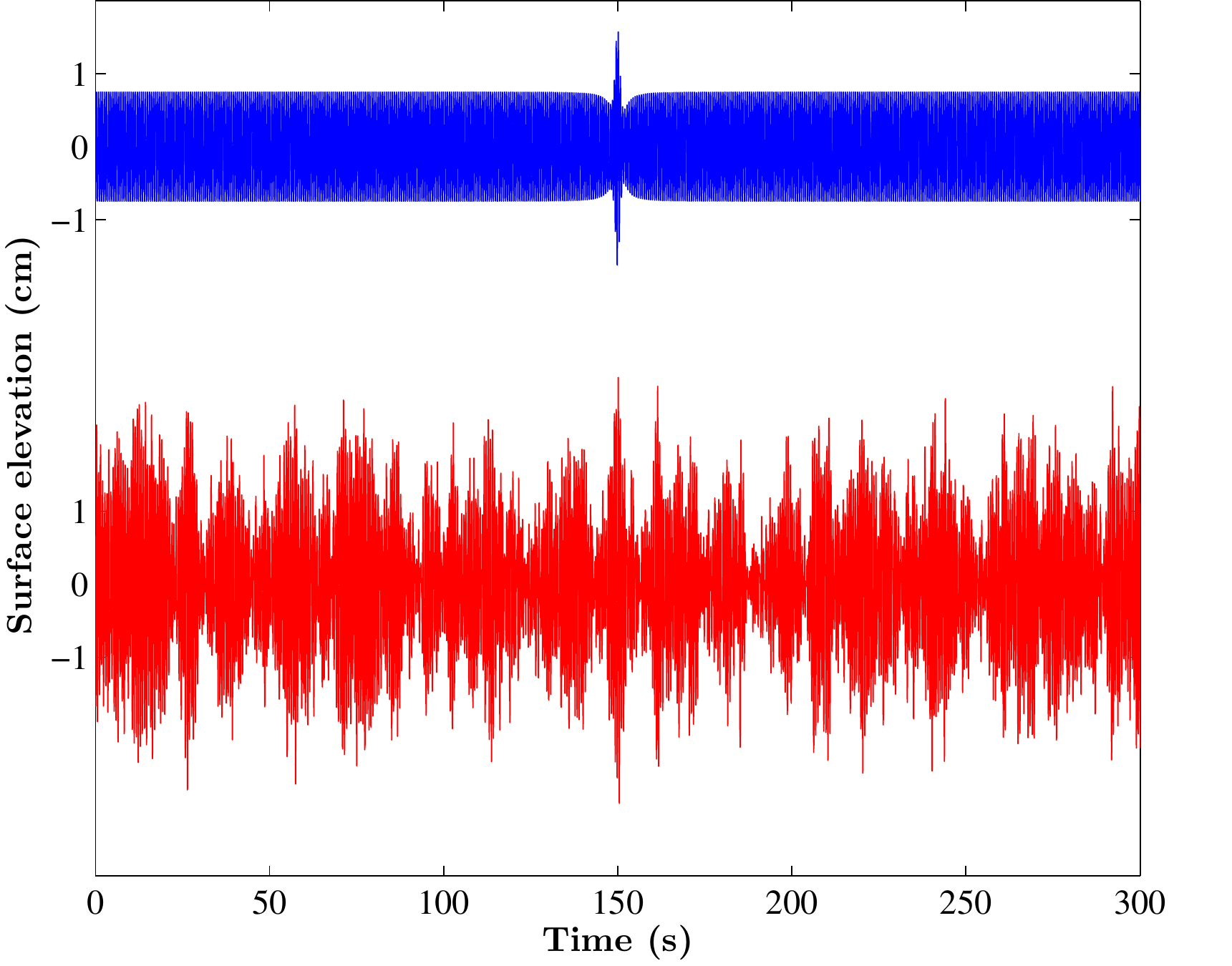}
\caption{(Color online) Upper panel: The Peregrine water surface displacement for $ak=0.08$ and $a=0.75$ cm, determined at $x=-6$ m (blue line). Lower Panel: The surface displacement of the PB of the upper panel, embedded in a JONSWAP wave field having a significant wave height of 3 cm with peak enhancement factor $\gamma=6$ at the same frequency of $f_p=1.7$ Hz (red line). This signal will then be used to generate the wave motion by the wave maker.}\label{fig2}
\end{figure}

The experiments have been performed in a deep-water facility, see details in \cite{kibler2015superregular}. The dimensional amplitude of the carrier has been set to be $a=0.75$ cm, while the the steepness is $ak=0.08$. Thus, the wave peak frequency is $f_p=1.7$ Hz. Considering the expression of the water surface elevation being
\begin{equation}
\eta\left(x,t\right)=\operatorname{Re}\left(\Psi\left(x,t\right)\exp[i\left(kx-\omega t\right)]\right), 
\end{equation}
the temporal surface displacement of the Peregrine model $\eta_P(t)$ is determined in the expectation to observe the theoretical maximal breather compression 6 m from the wave generator, i.e. $\eta_P(t)=\eta_P(x=-6,t)$, see upper panel of Fig. \ref{fig2}.  
In the next step $\eta_P(t)$ will be embedded in a chaotic wave field. Generally, one possibility to generate realistic oceanic sea states, is for the energy of the irregular wave field to satisfy a JONSWAP spectrum \cite{komen1996dynamics}
\begin{eqnarray} 
S(f)=\dfrac{\alpha}{f^5}\exp\left[-\dfrac{5}{4}\left(\dfrac{f_p}{f}\right)^4\right]\gamma^{\exp\left[-\dfrac{\left(f-f_p\right)^2}{2\sigma^2f_p^2}\right]}.
\end{eqnarray}
We set the frequency peak at $f_p=1.7$ Hz, the significant wave height of the wave field, defined as four times the standard deviation of the wave field \cite{kharif2009rogue}, to be $H_s=3$ cm and the enhancement factor $\gamma=6$. Furthermore, $\sigma=0.09$ if $f>f_p$ and $\sigma=0.07$ if $f\leq f_p$. Note that the JONSWAP spectrum is just a peaked-enhanced extension of the Pierson-Moskowitz spectrum \cite{komen1996dynamics}. A JONSWAP surface displacement realization with random phases $\varphi_n\in]0,2\pi[$ is then determined by \cite{onorato2001freak}
\begin{eqnarray}
\eta_{\textnormal{\tiny JONSWAP}}(0,t)&=\sum\limits_{i=1}^{N}{\sqrt{2S\left(f_n\right)\Delta f_n}\cos\left(2\pi f_n t-\varphi_n\right)}.\nonumber\\
&
\end{eqnarray} 
The Peregrine surface elevation $\eta_P(t)$ is now added to a JONSWAP realization with random phases $\eta_R(t)$, as described above, in which the main peak of the latter has now been removed accordingly to be replaced by the Peregrine energy in the new constructed hybrid surface elevation
\begin{eqnarray}
\eta_{\textnormal{hybrid}}(t)=\eta_P(t)+\eta_R(t)
\label{seh}
\end{eqnarray} 
The hybrid time-series (\ref{seh}) is therefore a JONSWAP realization, with parameters as mentioned above, having a Peregrine energy peak. It is shown in the lower panel of Fig. \ref{fig2} and is now chosen as a boundary condition to drive the wave maker. We point out that the unstable Peregrine envelope perturbation is now cloaked in the JONSWAP wave train, which reveals several in wave height similar wave modulations as the Peregrine-type wave packet. Note that MI in JONSWAP random sea states has been discussed in a general context for instance theoretically within the framework of the NLSE and the inverse scattering transform (IST) in \cite{islas2005predicting,osborne2010nonlinear}, while numerically in \cite{onorato2001freak,toenger2015emergent} and experimentally in \cite{onorato2006extreme}.

\begin{figure}[h]
\centering
\includegraphics[width=\columnwidth]{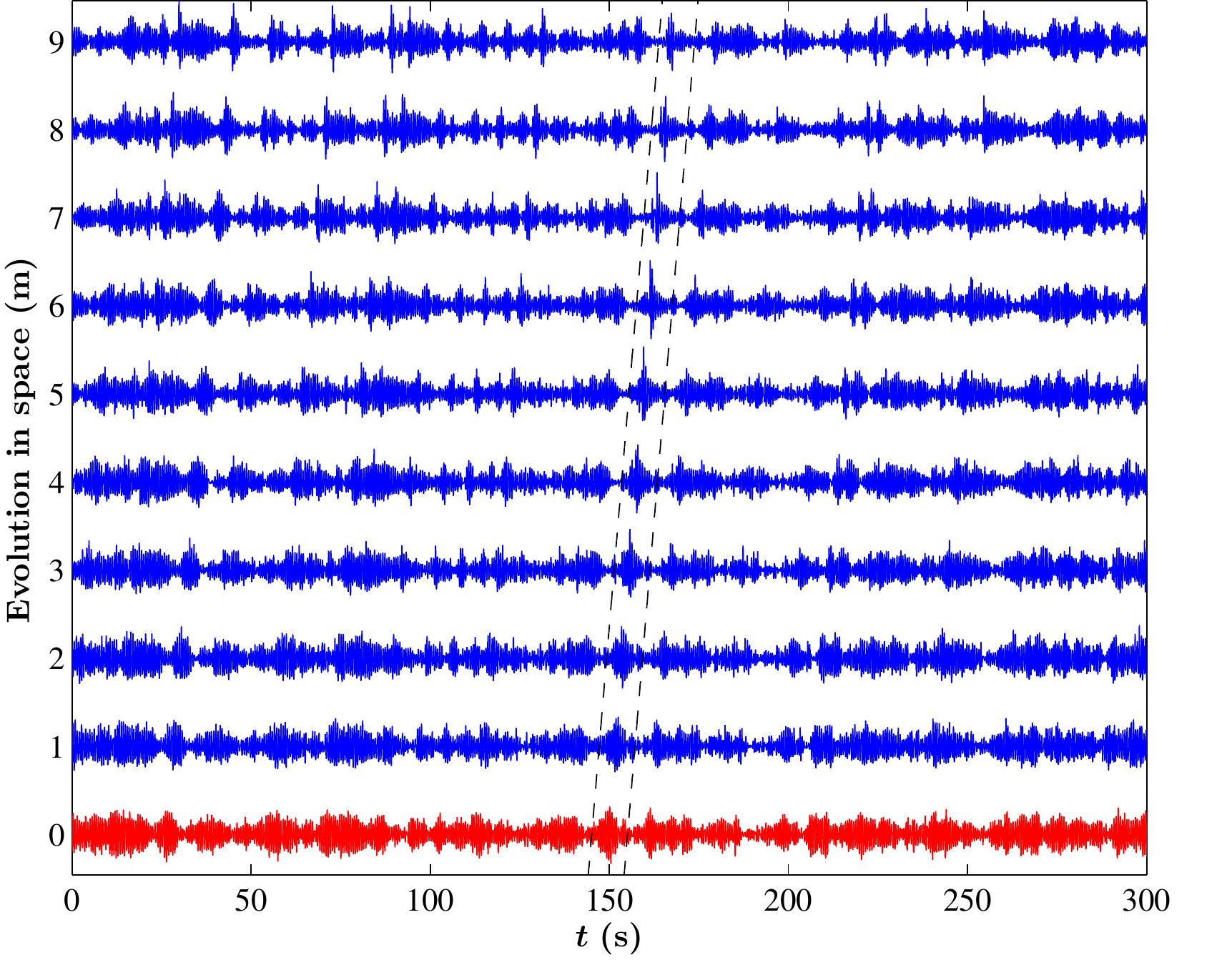}
\caption{(Color online) Evolution of the hybrid JONSWAP-Peregrine wave field over 9 m. The bottom time-series (red line) illustrates the boundary conditions, while the dashed lines limit the unstable Peregrine packet, moving with the group velocity $c_g=\dfrac{\omega}{2k}=0.46$ m$\cdot$s$^{-1}$.}\label{fig3}
\end{figure}

The evolution of the generated wave field is measured equidistantly at nine positions along the wave flume. The last wave gauge is placed 9 m from the wave maker, that is still at 3 m distance from the beach, hence, far enough to be affected by strong wave reflections. Fig. \ref{fig3} depicts the propagation of the wave field with particular emphasis on the hybrid Peregrine packets, in the intervals bounded by the dashed lines.

\begin{figure}[h]
\centering
\includegraphics[width=\columnwidth]{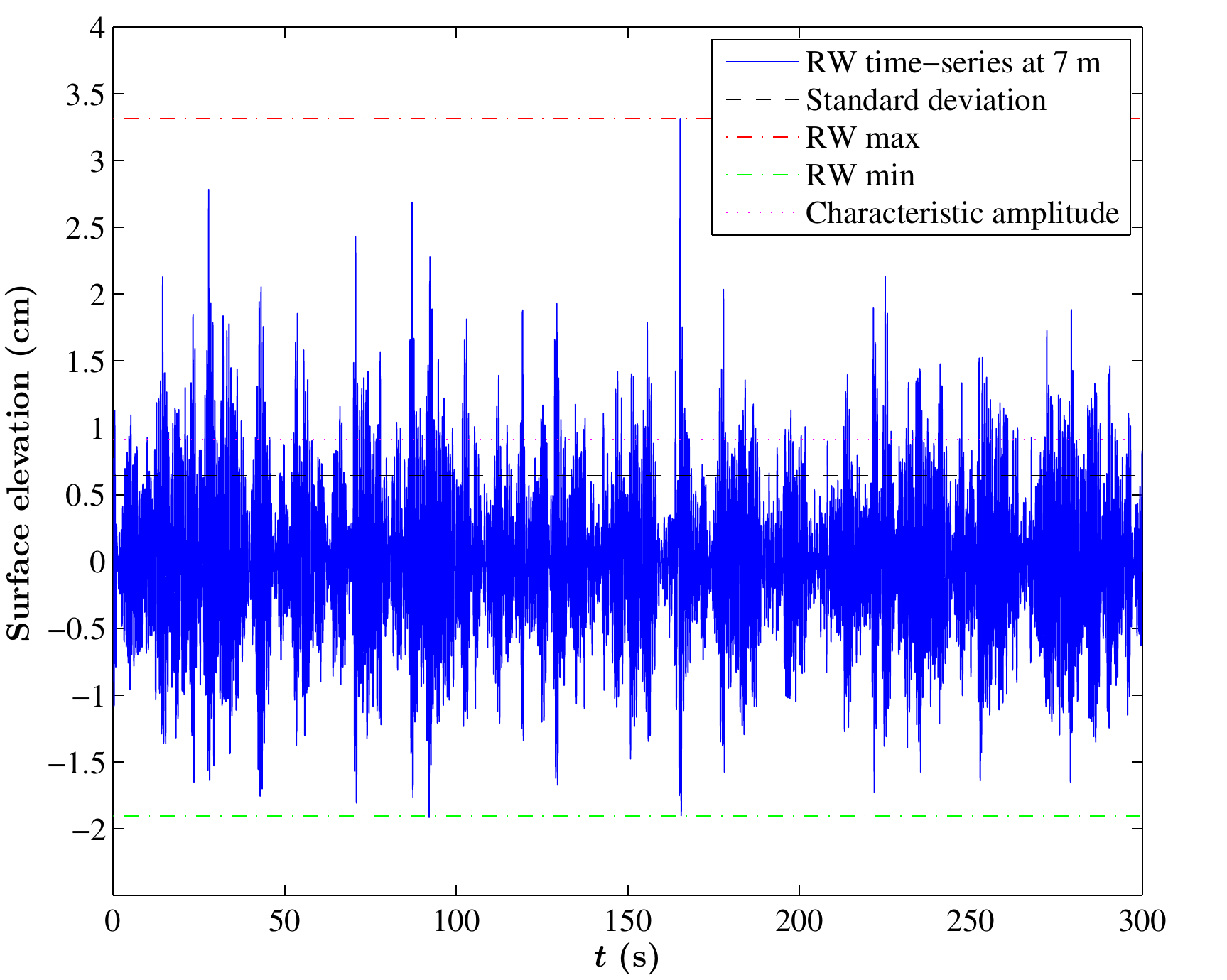}
\caption{(Color online) Temporal wave displacement of the maximal Peregrine-type rogue waves, measured 7 m from the wave maker. The horizontal lines are described in the Fig.'s caption.}\label{fig4}
\end{figure}

Indeed, we can clearly notice a significant focusing during the propagation of a Peregrine-type wave packet evolving in the irregular water wave field. The maximal wave is measured at 7 m and highlights a wave height of 5.22 cm. This latter wave train is isolated and shown seperately in Fig. \ref{fig4}. 
As a matter of fact, we can state that this Peregrine-type extreme wave is a rogue wave, since the abnormality index of the maximal wave, defined of being the ratio of maximal wave height and significant wave height, exceeds two. Other physical features of this extreme wave are summarized in Table \ref{tab1}. 

It is emphasized that due to strong focusing of the wave, slight spilling breaking has been observed during the evolution. Nevertheless, these measurements prove that Peregrine dynamics may indeed persist in a random one dimensional sea states with strong irregularities, allowing therefore tracking extreme oceanic event to backbone models of integrable evolution equations. This also justifies once again the choice of investigating fundamental theoretical as well as physical properties of exact solutions in order to accurately predict RWs in the ocean \cite{kharif2009rogue,osborne2010nonlinear}. 

\begin{center}
\begin{figure}[h]
\centering
\begin{tabular}{c}
\includegraphics[width=.4\textwidth]{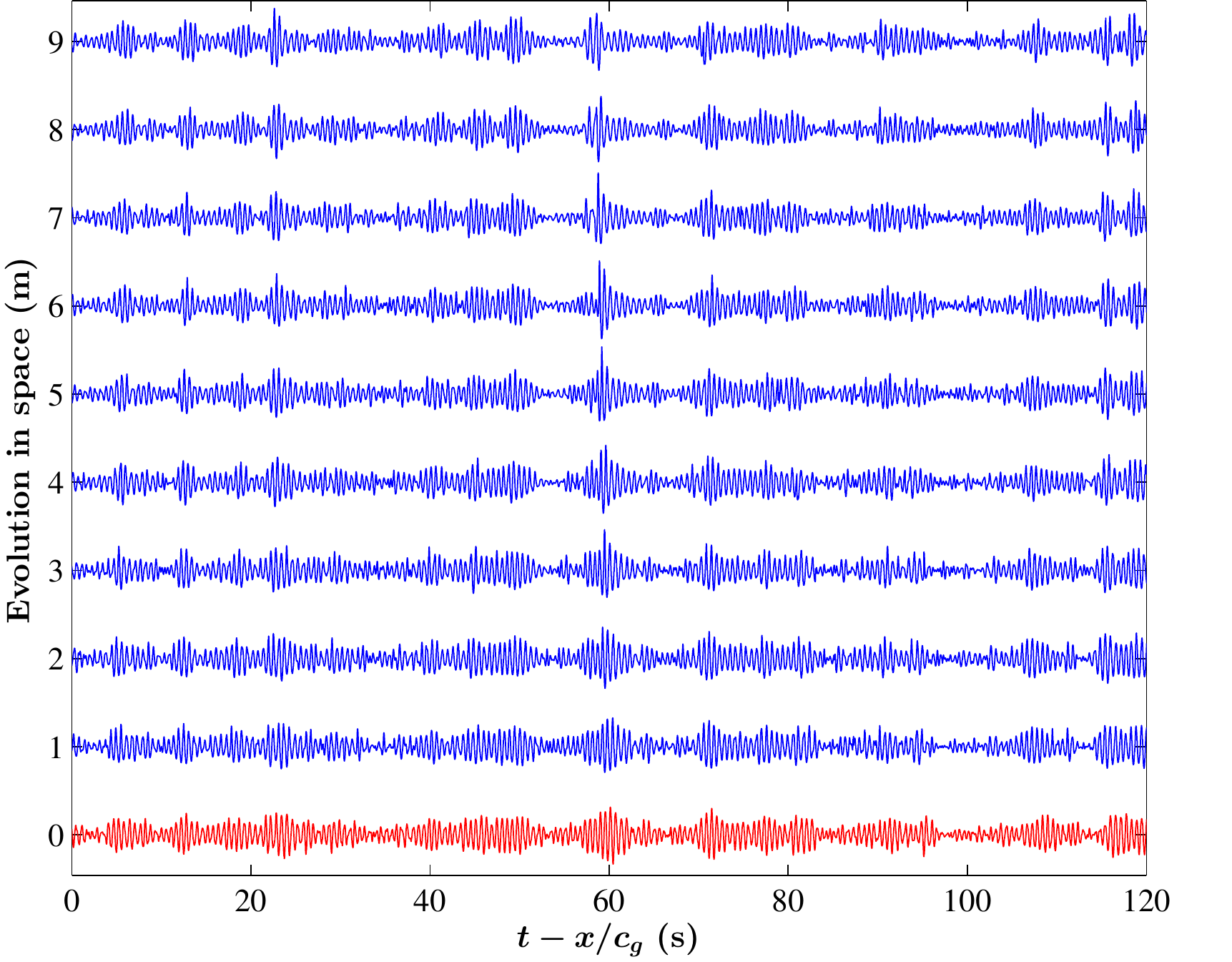}\\
\end{tabular}
\begin{tabular}{cc}
\includegraphics[width=.49\textwidth]{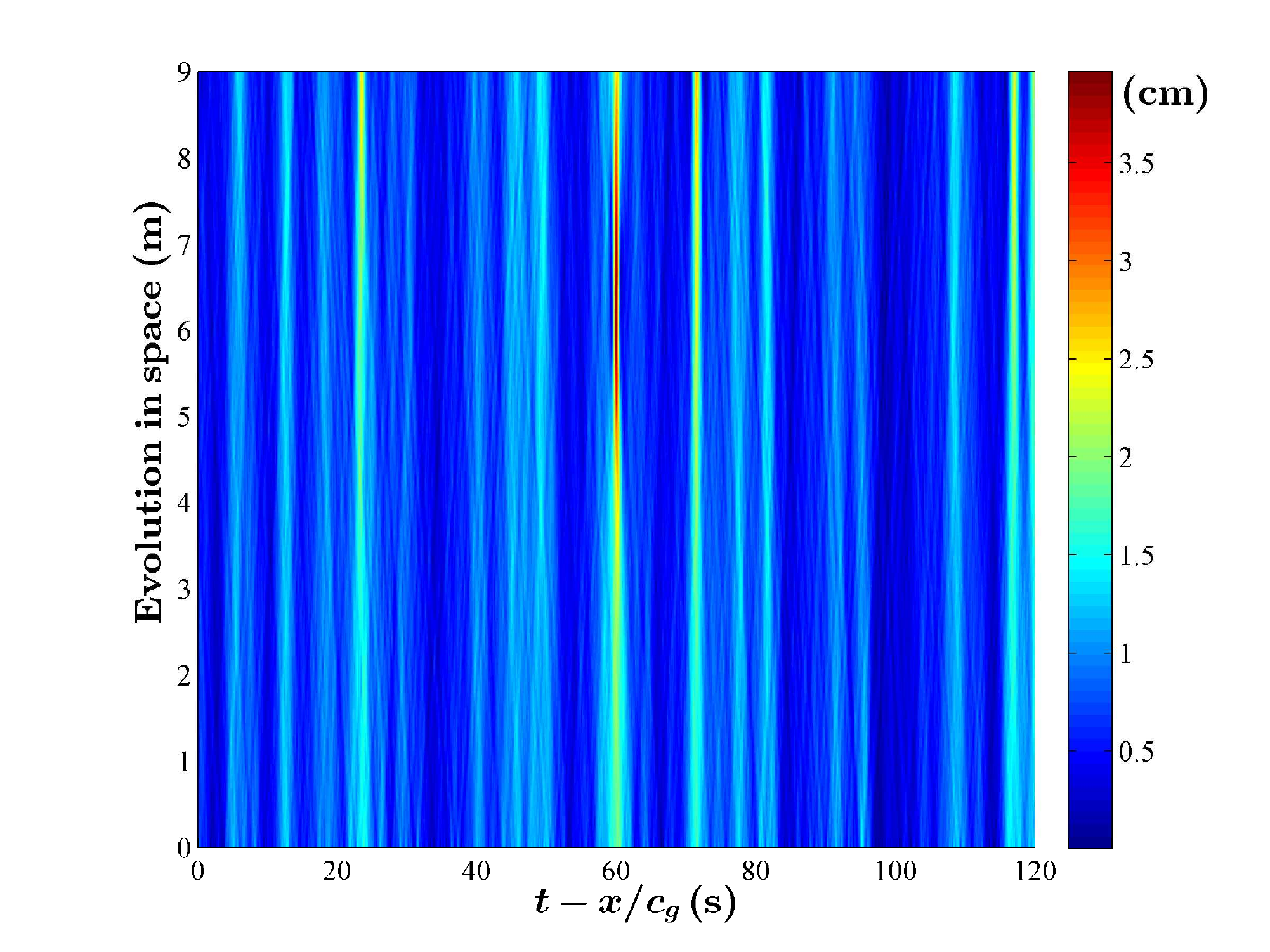}&
\includegraphics[width=.49\textwidth]{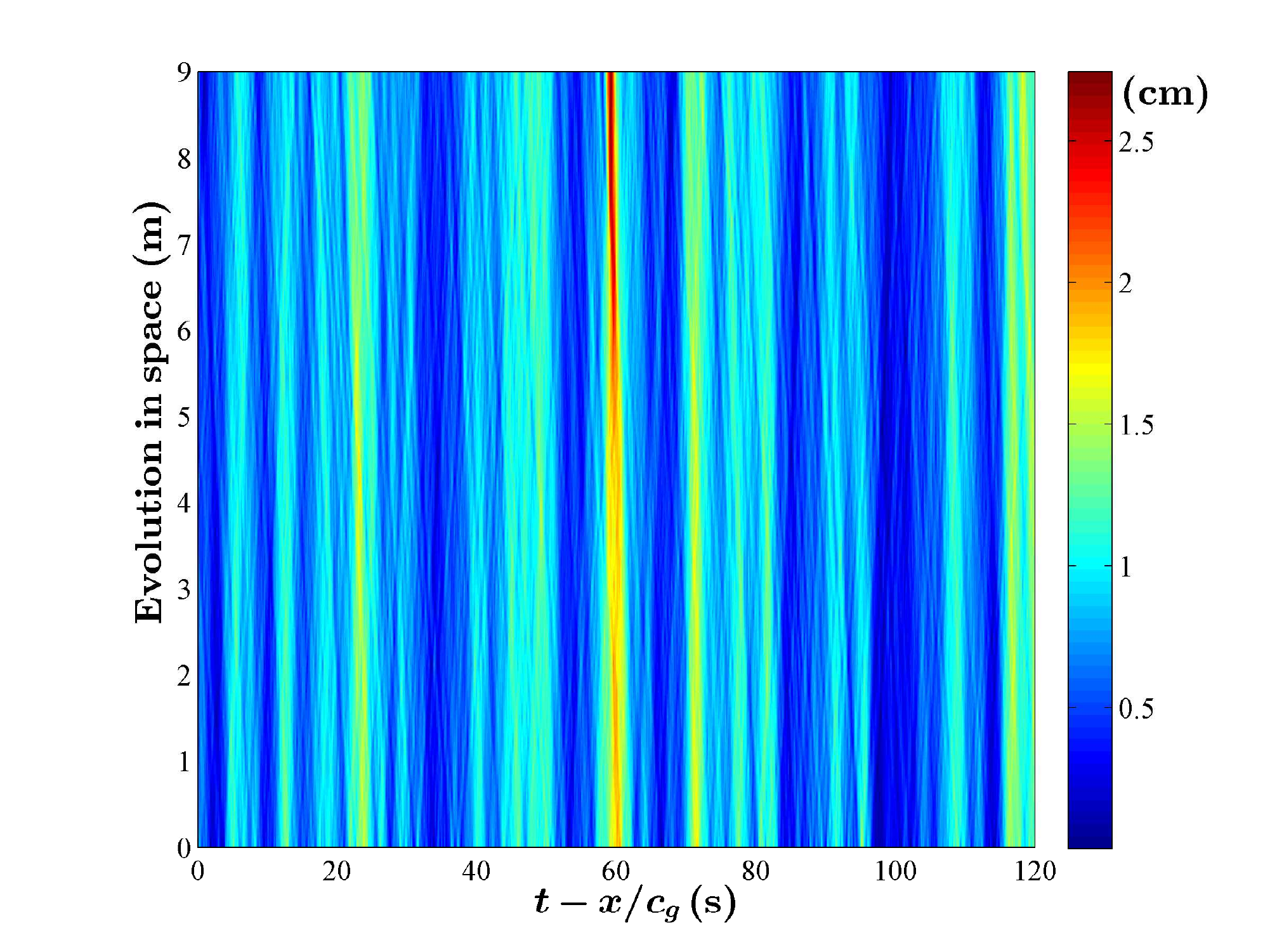}\\
\end{tabular}
\caption{(Color online) Upper panel: Surface displacements aligned by the value of the group velocity $c_g$. Lower left panel: NLSE simulation prediction results, starting from the same initial condition as in the laboratory experiments. Lower right panel: MNLSE simulation prediction results, also starting from the same initial condition as in the laboratory experiments.}
\label{fig5}
\end{figure}
\end{center}

As next, the experimental wave evolution is compared to numerical simulations, based on NLSE and MNLSE, using the split-step method. The time-MNLSE \cite{dysthe1979note,trulsen2001spatial} reads
\begin{eqnarray} 
&\operatorname{i}\left(\Psi_x+\dfrac{2k}{\omega}\Psi_t\right)-\dfrac{1}{g}\Psi_{tt}-k^3\left|\Psi\right|^2\Psi\nonumber\\
&-\operatorname{i}\dfrac{k^3}{\omega}\left(6\left|\Psi\right|^2\dfrac{\partial\Psi}{\partial t}+2\Psi\dfrac{\partial\left|\Psi\right|^2}{\partial t}-2\operatorname{i}\Psi\mathcal{H}\left[\dfrac{\partial\left|\Psi\right|^2}{\partial t}\right]\right)=0\nonumber\\,
\label{MNLS}
\end{eqnarray} 
while $\mathcal{H}$ denotes the Hilbert transform. The MNLSE is an extension of the NLSE that improves approximation of dispersion and that takes into account the mean flow of the wave field. The surface measurement, restricted to 120 s and aligned with respect to the group velocity $c_g$ as well as both simulation results are illustrated in Fig. \ref{fig5}. 
\onecolumngrid 
\begin{center}
\begin{table}[h]
\begin{tabular}{|c|c|c|c|c|c|}
\hline
Standard deviation &  Characteristic amplitude & Significant wave height & Maximal height & Abnormality index \\ 
\hline $\sigma=0.64$ cm &  $a_{\textnormal{char}}=0.90$ cm & $H_s=2.56$ cm & $H_{\textnormal{max}}=5.22$ cm & $AI$=2.04 \\ \hline
\end{tabular}
\caption{Characteristic properties of the maximal wave in the 300 s wave train, measured 7 m from the wave generator.} \label{tab1}
\end{table} 
\end{center}
\twocolumngrid 
Both simulation results are in qualitative good agreement with the laboratory experiments. We recall that neither the NLS nor the MNLSE can model the breaking of the wave field since at this stage of approximation the wave field is assumed to be irrotational, among other limitations. In fact, both evolution equations prove to be very accurate in predicting the extreme event, occurring roughly at the expected distance from the wave generator, while surrounding modulated wave packets remain stable during their evolution. Furthermore, it is interesting to notice that the maximal wave amplification in the MNLSE prediction occurs later compared to the NLSE, in agreement with theory. Even though being generally less accurate than the MNLSE when the wave process becomes broad-banded \cite{trulsen1996modified,chabchoub2013hydrodynamic}, the NLSE simulations surprsingly provide a better estimate to the start of growth and decay of wave compression in the experiment. These simulations also vindicate the application of these evolution equations for ocean waves \cite{trulsen2007weakly}.

To conclude, we have shown that doubly-localized PB dynamics may persist on irregular background. Indeed, the constructed hybrid Peregrine-JONSWAP wave field with random phases is shown to generate a hydrodynamic extreme event at the expected temporal and spatial locality. The observed highest wave has an abnormality index that exceeds two, satisfying the definition of ocean RWs. The experimental results are effectively in good agreement with NLSE and MNLSE simulations, both accurate in the prediction of the single extreme event despite constraints in the modeling that does not include viscosity, dissipation, breaking, and other limitations associated to laboratory experiments. Future work will characterize the influence of initial JONSWAP amplitudes, mode phases as well as spectral parameters $\alpha$ and $\gamma$, which have a significant influence when interacting with NLSE breathers in the described approach. Numerical simulations based on more accurate evolution equations, such as the higher-order spectral method \cite{ducrozet2012modified,slunyaev2013highest}, may characterize the limitations of the approach as well as reveal new insights to the problem, taking into account that the latter are much faster to perform compared to laboratory experiments. This study also discloses that characteristic breather spectral properties in physical domain \cite{akhmediev2011roguespectra} as well as in the IST plane \cite{osborne2010nonlinear,randoux2015identification} are indeed promising features that can be applied for the sake of accurate deterministic extreme event detection. Due to the interdisciplinary character of the approach it is expected that analogous numerical and experimental studies may be motivated for instance in Kerr media and plasma, hence, improving the decryption of RWs' as well as nonlinear localized wave motions' in regular and irregular states.

A.C. acknowledges the Japan Society for the Promotion of Science (JSPS), The Association of German Engineers (VDI) and the Burgundy Region.


\bibliography{RefAC}

\begin{thebibliography}{37}%
\makeatletter
\providecommand \@ifxundefined [1]{%
 \@ifx{#1\undefined}
}%
\providecommand \@ifnum [1]{%
 \ifnum #1\expandafter \@firstoftwo
 \else \expandafter \@secondoftwo
 \fi
}%
\providecommand \@ifx [1]{%
 \ifx #1\expandafter \@firstoftwo
 \else \expandafter \@secondoftwo
 \fi
}%
\providecommand \natexlab [1]{#1}%
\providecommand \enquote  [1]{``#1''}%
\providecommand \bibnamefont  [1]{#1}%
\providecommand \bibfnamefont [1]{#1}%
\providecommand \citenamefont [1]{#1}%
\providecommand \href@noop [0]{\@secondoftwo}%
\providecommand \href [0]{\begingroup \@sanitize@url \@href}%
\providecommand \@href[1]{\@@startlink{#1}\@@href}%
\providecommand \@@href[1]{\endgroup#1\@@endlink}%
\providecommand \@sanitize@url [0]{\catcode `\\12\catcode `\$12\catcode
  `\&12\catcode `\#12\catcode `\^12\catcode `\_12\catcode `\%12\relax}%
\providecommand \@@startlink[1]{}%
\providecommand \@@endlink[0]{}%
\providecommand \url  [0]{\begingroup\@sanitize@url \@url }%
\providecommand \@url [1]{\endgroup\@href {#1}{\urlprefix }}%
\providecommand \urlprefix  [0]{URL }%
\providecommand \Eprint [0]{\href }%
\providecommand \doibase [0]{http://dx.doi.org/}%
\providecommand \selectlanguage [0]{\@gobble}%
\providecommand \bibinfo  [0]{\@secondoftwo}%
\providecommand \bibfield  [0]{\@secondoftwo}%
\providecommand \translation [1]{[#1]}%
\providecommand \BibitemOpen [0]{}%
\providecommand \bibitemStop [0]{}%
\providecommand \bibitemNoStop [0]{.\EOS\space}%
\providecommand \EOS [0]{\spacefactor3000\relax}%
\providecommand \BibitemShut  [1]{\csname bibitem#1\endcsname}%
\let\auto@bib@innerbib\@empty
\bibitem [{\citenamefont {Kharif}\ \emph {et~al.}(2009)\citenamefont {Kharif},
  \citenamefont {Pelinovsky},\ and\ \citenamefont
  {Slunyaev}}]{kharif2009rogue}%
  \BibitemOpen
  \bibfield  {author} {\bibinfo {author} {\bibfnamefont {C.}~\bibnamefont
  {Kharif}}, \bibinfo {author} {\bibfnamefont {E.}~\bibnamefont {Pelinovsky}},
  \ and\ \bibinfo {author} {\bibfnamefont {A.}~\bibnamefont {Slunyaev}},\
  }\href@noop {} {\emph {\bibinfo {title} {Rogue waves in the ocean}}}\
  (\bibinfo  {publisher} {Springer},\ \bibinfo {year} {2009})\BibitemShut
  {NoStop}%
\bibitem [{\citenamefont {Osborne}(2010)}]{osborne2010nonlinear}%
  \BibitemOpen
  \bibfield  {author} {\bibinfo {author} {\bibfnamefont {A.}~\bibnamefont
  {Osborne}},\ }\href@noop {} {\emph {\bibinfo {title} {Nonlinear Ocean Waves
  \& the Inverse Scattering Transform}}},\ Vol.~\bibinfo {volume} {97}\
  (\bibinfo  {publisher} {Academic Press},\ \bibinfo {year} {2010})\BibitemShut
  {NoStop}%
\bibitem [{\citenamefont {Benjamin}\ and\ \citenamefont
  {Feir}(1967)}]{benjamin1967disintegration}%
  \BibitemOpen
  \bibfield  {author} {\bibinfo {author} {\bibfnamefont {T.~B.}\ \bibnamefont
  {Benjamin}}\ and\ \bibinfo {author} {\bibfnamefont {J.}~\bibnamefont
  {Feir}},\ }\href@noop {} {\bibfield  {journal} {\bibinfo  {journal} {J. Fluid
  Mech.}\ }\textbf {\bibinfo {volume} {27}},\ \bibinfo {pages} {417} (\bibinfo
  {year} {1967})}\BibitemShut {NoStop}%
\bibitem [{\citenamefont {Akhmediev}\ \emph {et~al.}(1985)\citenamefont
  {Akhmediev}, \citenamefont {Eleonskii},\ and\ \citenamefont
  {Kulagin}}]{akhmediev1985generation}%
  \BibitemOpen
  \bibfield  {author} {\bibinfo {author} {\bibfnamefont {N.}~\bibnamefont
  {Akhmediev}}, \bibinfo {author} {\bibfnamefont {V.~M.}\ \bibnamefont
  {Eleonskii}}, \ and\ \bibinfo {author} {\bibfnamefont {N.~E.}\ \bibnamefont
  {Kulagin}},\ }\href@noop {} {\bibfield  {journal} {\bibinfo  {journal} {Sov.
  Phys. JETP}\ }\textbf {\bibinfo {volume} {62}},\ \bibinfo {pages} {894 }
  (\bibinfo {year} {1985})}\BibitemShut {NoStop}%
\bibitem [{\citenamefont {Zakharov}\ and\ \citenamefont
  {Gelash}(2013)}]{zakharov2013nonlinear}%
  \BibitemOpen
  \bibfield  {author} {\bibinfo {author} {\bibfnamefont {V.}~\bibnamefont
  {Zakharov}}\ and\ \bibinfo {author} {\bibfnamefont {A.}~\bibnamefont
  {Gelash}},\ }\href@noop {} {\bibfield  {journal} {\bibinfo  {journal}
  {Physical review letters}\ }\textbf {\bibinfo {volume} {111}},\ \bibinfo
  {pages} {054101} (\bibinfo {year} {2013})}\BibitemShut {NoStop}%
\bibitem [{\citenamefont {Kharif}\ and\ \citenamefont
  {Pelinovsky}(2003)}]{kharif2003physical}%
  \BibitemOpen
  \bibfield  {author} {\bibinfo {author} {\bibfnamefont {C.}~\bibnamefont
  {Kharif}}\ and\ \bibinfo {author} {\bibfnamefont {E.}~\bibnamefont
  {Pelinovsky}},\ }\href@noop {} {\bibfield  {journal} {\bibinfo  {journal}
  {European J. Mechanics-B/Fluids}\ }\textbf {\bibinfo {volume} {22}},\
  \bibinfo {pages} {603} (\bibinfo {year} {2003})}\BibitemShut {NoStop}%
\bibitem [{\citenamefont {Onorato}\ \emph
  {et~al.}(2013{\natexlab{a}})\citenamefont {Onorato}, \citenamefont
  {Residori}, \citenamefont {Bortolozzo}, \citenamefont {Montina},\ and\
  \citenamefont {Arecchi}}]{onorato2013roguereport}%
  \BibitemOpen
  \bibfield  {author} {\bibinfo {author} {\bibfnamefont {M.}~\bibnamefont
  {Onorato}}, \bibinfo {author} {\bibfnamefont {S.}~\bibnamefont {Residori}},
  \bibinfo {author} {\bibfnamefont {U.}~\bibnamefont {Bortolozzo}}, \bibinfo
  {author} {\bibfnamefont {A.}~\bibnamefont {Montina}}, \ and\ \bibinfo
  {author} {\bibfnamefont {F.~T.}\ \bibnamefont {Arecchi}},\ }\href@noop {}
  {\bibfield  {journal} {\bibinfo  {journal} {Physics Reports}\ }\textbf
  {\bibinfo {volume} {528}},\ \bibinfo {pages} {47} (\bibinfo {year}
  {2013}{\natexlab{a}})}\BibitemShut {NoStop}%
\bibitem [{\citenamefont {Dudley}\ \emph {et~al.}(2014)\citenamefont {Dudley},
  \citenamefont {Dias}, \citenamefont {Erkintalo},\ and\ \citenamefont
  {Genty}}]{dudley2014instabilities}%
  \BibitemOpen
  \bibfield  {author} {\bibinfo {author} {\bibfnamefont {J.~M.}\ \bibnamefont
  {Dudley}}, \bibinfo {author} {\bibfnamefont {F.}~\bibnamefont {Dias}},
  \bibinfo {author} {\bibfnamefont {M.}~\bibnamefont {Erkintalo}}, \ and\
  \bibinfo {author} {\bibfnamefont {G.}~\bibnamefont {Genty}},\ }\href@noop {}
  {\bibfield  {journal} {\bibinfo  {journal} {Nature Photonics}\ }\textbf
  {\bibinfo {volume} {8}},\ \bibinfo {pages} {755} (\bibinfo {year}
  {2014})}\BibitemShut {NoStop}%
\bibitem [{\citenamefont {Baronio}\ \emph {et~al.}(2014)\citenamefont
  {Baronio}, \citenamefont {Conforti}, \citenamefont {Degasperis},
  \citenamefont {Lombardo}, \citenamefont {Onorato},\ and\ \citenamefont
  {Wabnitz}}]{baronio2014vector}%
  \BibitemOpen
  \bibfield  {author} {\bibinfo {author} {\bibfnamefont {F.}~\bibnamefont
  {Baronio}}, \bibinfo {author} {\bibfnamefont {M.}~\bibnamefont {Conforti}},
  \bibinfo {author} {\bibfnamefont {A.}~\bibnamefont {Degasperis}}, \bibinfo
  {author} {\bibfnamefont {S.}~\bibnamefont {Lombardo}}, \bibinfo {author}
  {\bibfnamefont {M.}~\bibnamefont {Onorato}}, \ and\ \bibinfo {author}
  {\bibfnamefont {S.}~\bibnamefont {Wabnitz}},\ }\href@noop {} {\bibfield
  {journal} {\bibinfo  {journal} {Phys. Rev. Lett.}\ }\textbf {\bibinfo
  {volume} {113}},\ \bibinfo {pages} {034101} (\bibinfo {year}
  {2014})}\BibitemShut {NoStop}%
\bibitem [{\citenamefont {Akhmediev}\ and\ \citenamefont
  {Ankiewicz}(1997)}]{akhmediev1997solitons}%
  \BibitemOpen
  \bibfield  {author} {\bibinfo {author} {\bibfnamefont {N.}~\bibnamefont
  {Akhmediev}}\ and\ \bibinfo {author} {\bibfnamefont {A.}~\bibnamefont
  {Ankiewicz}},\ }\href@noop {} {\emph {\bibinfo {title} {{Solitons: Nonlinear
  pulses and beams}}}}\ (\bibinfo  {publisher} {Chapman \& Hall},\ \bibinfo
  {year} {1997})\BibitemShut {NoStop}%
\bibitem [{\citenamefont {Peregrine}(1983)}]{peregrine1983water}%
  \BibitemOpen
  \bibfield  {author} {\bibinfo {author} {\bibfnamefont {D.~H.}\ \bibnamefont
  {Peregrine}},\ }\href@noop {} {\bibfield  {journal} {\bibinfo  {journal} {J.
  Australian Math. Soc. Series B. Applied Mathematics}\ }\textbf {\bibinfo
  {volume} {25}},\ \bibinfo {pages} {16} (\bibinfo {year} {1983})}\BibitemShut
  {NoStop}%
\bibitem [{\citenamefont {Shrira}\ and\ \citenamefont
  {Geogjaev}(2010)}]{shrira2010makes}%
  \BibitemOpen
  \bibfield  {author} {\bibinfo {author} {\bibfnamefont {V.~I.}\ \bibnamefont
  {Shrira}}\ and\ \bibinfo {author} {\bibfnamefont {V.~V.}\ \bibnamefont
  {Geogjaev}},\ }\href@noop {} {\bibfield  {journal} {\bibinfo  {journal} {J.
  Eng. Math.}\ }\textbf {\bibinfo {volume} {67}},\ \bibinfo {pages} {11}
  (\bibinfo {year} {2010})}\BibitemShut {NoStop}%
\bibitem [{\citenamefont {Kibler}\ \emph {et~al.}(2010)\citenamefont {Kibler},
  \citenamefont {Fatome}, \citenamefont {Finot}, \citenamefont {Millot},
  \citenamefont {Dias}, \citenamefont {Genty}, \citenamefont {Akhmediev},\ and\
  \citenamefont {Dudley}}]{kibler2010peregrine}%
  \BibitemOpen
  \bibfield  {author} {\bibinfo {author} {\bibfnamefont {B.}~\bibnamefont
  {Kibler}}, \bibinfo {author} {\bibfnamefont {J.}~\bibnamefont {Fatome}},
  \bibinfo {author} {\bibfnamefont {C.}~\bibnamefont {Finot}}, \bibinfo
  {author} {\bibfnamefont {G.}~\bibnamefont {Millot}}, \bibinfo {author}
  {\bibfnamefont {F.}~\bibnamefont {Dias}}, \bibinfo {author} {\bibfnamefont
  {G.}~\bibnamefont {Genty}}, \bibinfo {author} {\bibfnamefont
  {N.}~\bibnamefont {Akhmediev}}, \ and\ \bibinfo {author} {\bibfnamefont
  {J.~M.}\ \bibnamefont {Dudley}},\ }\href@noop {} {\bibfield  {journal}
  {\bibinfo  {journal} {Nature Physics}\ }\textbf {\bibinfo {volume} {6}},\
  \bibinfo {pages} {790} (\bibinfo {year} {2010})}\BibitemShut {NoStop}%
\bibitem [{\citenamefont {Chabchoub}\ \emph {et~al.}(2011)\citenamefont
  {Chabchoub}, \citenamefont {Hoffmann},\ and\ \citenamefont
  {Akhmediev}}]{chabchoub2011rogue}%
  \BibitemOpen
  \bibfield  {author} {\bibinfo {author} {\bibfnamefont {A.}~\bibnamefont
  {Chabchoub}}, \bibinfo {author} {\bibfnamefont {N.}~\bibnamefont {Hoffmann}},
  \ and\ \bibinfo {author} {\bibfnamefont {N.}~\bibnamefont {Akhmediev}},\
  }\href@noop {} {\bibfield  {journal} {\bibinfo  {journal} {Phys. Rev. Lett.}\
  }\textbf {\bibinfo {volume} {106}},\ \bibinfo {pages} {204502} (\bibinfo
  {year} {2011})}\BibitemShut {NoStop}%
\bibitem [{\citenamefont {Bailung}\ \emph {et~al.}(2011)\citenamefont
  {Bailung}, \citenamefont {Sharma},\ and\ \citenamefont
  {Nakamura}}]{bailung2011observation}%
  \BibitemOpen
  \bibfield  {author} {\bibinfo {author} {\bibfnamefont {H.}~\bibnamefont
  {Bailung}}, \bibinfo {author} {\bibfnamefont {S.}~\bibnamefont {Sharma}}, \
  and\ \bibinfo {author} {\bibfnamefont {Y.}~\bibnamefont {Nakamura}},\
  }\href@noop {} {\bibfield  {journal} {\bibinfo  {journal} {Phys. Rev. Lett.}\
  }\textbf {\bibinfo {volume} {107}},\ \bibinfo {pages} {255005} (\bibinfo
  {year} {2011})}\BibitemShut {NoStop}%
\bibitem [{\citenamefont {Chabchoub}\ \emph
  {et~al.}(2013{\natexlab{a}})\citenamefont {Chabchoub}, \citenamefont
  {Hoffmann}, \citenamefont {Branger}, \citenamefont {Kharif},\ and\
  \citenamefont {Akhmediev}}]{chabchoub2013experimentswind}%
  \BibitemOpen
  \bibfield  {author} {\bibinfo {author} {\bibfnamefont {A.}~\bibnamefont
  {Chabchoub}}, \bibinfo {author} {\bibfnamefont {N.}~\bibnamefont {Hoffmann}},
  \bibinfo {author} {\bibfnamefont {H.}~\bibnamefont {Branger}}, \bibinfo
  {author} {\bibfnamefont {C.}~\bibnamefont {Kharif}}, \ and\ \bibinfo {author}
  {\bibfnamefont {N.}~\bibnamefont {Akhmediev}},\ }\href@noop {} {\bibfield
  {journal} {\bibinfo  {journal} {Physics of Fluids (1994-present)}\ }\textbf
  {\bibinfo {volume} {25}},\ \bibinfo {pages} {101704} (\bibinfo {year}
  {2013}{\natexlab{a}})}\BibitemShut {NoStop}%
\bibitem [{\citenamefont {Babanin}(2011)}]{babanin2011breaking}%
  \BibitemOpen
  \bibfield  {author} {\bibinfo {author} {\bibfnamefont {A.}~\bibnamefont
  {Babanin}},\ }\href@noop {} {\emph {\bibinfo {title} {Breaking and
  dissipation of ocean surface waves}}}\ (\bibinfo  {publisher} {Cambridge
  University Press},\ \bibinfo {year} {2011})\BibitemShut {NoStop}%
\bibitem [{\citenamefont {Zakharov}(1968)}]{zakharov1968stability}%
  \BibitemOpen
  \bibfield  {author} {\bibinfo {author} {\bibfnamefont {V.~E.}\ \bibnamefont
  {Zakharov}},\ }\href@noop {} {\bibfield  {journal} {\bibinfo  {journal} {J.
  Appl. Mech. Techn. Phys.}\ }\textbf {\bibinfo {volume} {9}},\ \bibinfo
  {pages} {190} (\bibinfo {year} {1968})}\BibitemShut {NoStop}%
\bibitem [{\citenamefont {Onorato}\ \emph
  {et~al.}(2013{\natexlab{b}})\citenamefont {Onorato}, \citenamefont {Proment},
  \citenamefont {Clauss},\ and\ \citenamefont
  {Klein}}]{onorato2013rogueberlin}%
  \BibitemOpen
  \bibfield  {author} {\bibinfo {author} {\bibfnamefont {M.}~\bibnamefont
  {Onorato}}, \bibinfo {author} {\bibfnamefont {D.}~\bibnamefont {Proment}},
  \bibinfo {author} {\bibfnamefont {G.}~\bibnamefont {Clauss}}, \ and\ \bibinfo
  {author} {\bibfnamefont {M.}~\bibnamefont {Klein}},\ }\href@noop {}
  {\bibfield  {journal} {\bibinfo  {journal} {PLOS ONE}\ }\textbf {\bibinfo
  {volume} {8}},\ \bibinfo {pages} {e54629} (\bibinfo {year}
  {2013}{\natexlab{b}})}\BibitemShut {NoStop}%
\bibitem [{\citenamefont {Alberello}\ \emph {et~al.}(2016)\citenamefont
  {Alberello}, \citenamefont {Chabchoub}, \citenamefont {Babanin},
  \citenamefont {Monty}, \citenamefont {Elsnab}, \citenamefont {Lee},
  \citenamefont {Bitner-Gregersen},\ and\ \citenamefont
  {Toffoli}}]{alberello2016velocity}%
  \BibitemOpen
  \bibfield  {author} {\bibinfo {author} {\bibfnamefont {A.}~\bibnamefont
  {Alberello}}, \bibinfo {author} {\bibfnamefont {A.}~\bibnamefont
  {Chabchoub}}, \bibinfo {author} {\bibfnamefont {A.~V.}\ \bibnamefont
  {Babanin}}, \bibinfo {author} {\bibfnamefont {J.~M.}\ \bibnamefont {Monty}},
  \bibinfo {author} {\bibfnamefont {J.}~\bibnamefont {Elsnab}}, \bibinfo
  {author} {\bibfnamefont {J.~H.}\ \bibnamefont {Lee}}, \bibinfo {author}
  {\bibfnamefont {E.~M.}\ \bibnamefont {Bitner-Gregersen}}, \ and\ \bibinfo
  {author} {\bibfnamefont {A.}~\bibnamefont {Toffoli}},\ }\href@noop {}
  {\bibfield  {journal} {\bibinfo  {journal} {ASME 2016 30th International
  Conference on Ocean, Offshore and Arctic Engineering}\ ,\ \bibinfo {pages}
  {54481}} (\bibinfo {year} {2016})}\BibitemShut {NoStop}%
\bibitem [{\citenamefont {Akhmediev}\ \emph {et~al.}(2009)\citenamefont
  {Akhmediev}, \citenamefont {Ankiewicz},\ and\ \citenamefont
  {Soto-Crespo}}]{akhmediev2009rogue}%
  \BibitemOpen
  \bibfield  {author} {\bibinfo {author} {\bibfnamefont {N.}~\bibnamefont
  {Akhmediev}}, \bibinfo {author} {\bibfnamefont {A.}~\bibnamefont
  {Ankiewicz}}, \ and\ \bibinfo {author} {\bibfnamefont {J.~M.}\ \bibnamefont
  {Soto-Crespo}},\ }\href@noop {} {\bibfield  {journal} {\bibinfo  {journal}
  {Phys. Rev. E}\ }\textbf {\bibinfo {volume} {80}},\ \bibinfo {pages} {026601}
  (\bibinfo {year} {2009})}\BibitemShut {NoStop}%
\bibitem [{\citenamefont {Chabchoub}\ \emph {et~al.}(2016)\citenamefont
  {Chabchoub}, \citenamefont {Onorato},\ and\ \citenamefont
  {Akhmediev}}]{chabchoub2016hydrodynamic}%
  \BibitemOpen
  \bibfield  {author} {\bibinfo {author} {\bibfnamefont {A.}~\bibnamefont
  {Chabchoub}}, \bibinfo {author} {\bibfnamefont {M.}~\bibnamefont {Onorato}},
  \ and\ \bibinfo {author} {\bibfnamefont {N.}~\bibnamefont {Akhmediev}},\
  }\href@noop {} {\bibfield  {journal} {\bibinfo  {journal} {Rogue and Shock
  Waves, M. Onorato (Editor), Lecture Notes in Physics, Springer}\ } (\bibinfo
  {year} {2016})}\BibitemShut {NoStop}%
\bibitem [{\citenamefont {Kibler}\ \emph {et~al.}(2015)\citenamefont {Kibler},
  \citenamefont {Chabchoub}, \citenamefont {Gelash}, \citenamefont
  {Akhmediev},\ and\ \citenamefont {Zakharov}}]{kibler2015superregular}%
  \BibitemOpen
  \bibfield  {author} {\bibinfo {author} {\bibfnamefont {B.}~\bibnamefont
  {Kibler}}, \bibinfo {author} {\bibfnamefont {A.}~\bibnamefont {Chabchoub}},
  \bibinfo {author} {\bibfnamefont {A.}~\bibnamefont {Gelash}}, \bibinfo
  {author} {\bibfnamefont {N.}~\bibnamefont {Akhmediev}}, \ and\ \bibinfo
  {author} {\bibfnamefont {V.~E.}\ \bibnamefont {Zakharov}},\ }\href@noop {}
  {\bibfield  {journal} {\bibinfo  {journal} {Phys. Rev. X}\ }\textbf {\bibinfo
  {volume} {5}},\ \bibinfo {pages} {041026} (\bibinfo {year}
  {2015})}\BibitemShut {NoStop}%
\bibitem [{\citenamefont {Komen}\ \emph {et~al.}(1996)\citenamefont {Komen},
  \citenamefont {Cavaleri}, \citenamefont {Donelan}, \citenamefont
  {Hasselmann}, \citenamefont {Hasselmann},\ and\ \citenamefont
  {Janssen}}]{komen1996dynamics}%
  \BibitemOpen
  \bibfield  {author} {\bibinfo {author} {\bibfnamefont {G.~J.}\ \bibnamefont
  {Komen}}, \bibinfo {author} {\bibfnamefont {L.}~\bibnamefont {Cavaleri}},
  \bibinfo {author} {\bibfnamefont {M.}~\bibnamefont {Donelan}}, \bibinfo
  {author} {\bibfnamefont {K.}~\bibnamefont {Hasselmann}}, \bibinfo {author}
  {\bibfnamefont {S.}~\bibnamefont {Hasselmann}}, \ and\ \bibinfo {author}
  {\bibfnamefont {P.}~\bibnamefont {Janssen}},\ }\href@noop {} {\emph {\bibinfo
  {title} {Dynamics and modelling of ocean waves}}}\ (\bibinfo  {publisher}
  {Cambridge university press},\ \bibinfo {year} {1996})\BibitemShut {NoStop}%
\bibitem [{\citenamefont {Onorato}\ \emph {et~al.}(2001)\citenamefont
  {Onorato}, \citenamefont {Osborne}, \citenamefont {Serio},\ and\
  \citenamefont {Bertone}}]{onorato2001freak}%
  \BibitemOpen
  \bibfield  {author} {\bibinfo {author} {\bibfnamefont {M.}~\bibnamefont
  {Onorato}}, \bibinfo {author} {\bibfnamefont {A.~R.}\ \bibnamefont
  {Osborne}}, \bibinfo {author} {\bibfnamefont {M.}~\bibnamefont {Serio}}, \
  and\ \bibinfo {author} {\bibfnamefont {S.}~\bibnamefont {Bertone}},\
  }\href@noop {} {\bibfield  {journal} {\bibinfo  {journal} {Physical Review
  Letters}\ }\textbf {\bibinfo {volume} {86}},\ \bibinfo {pages} {5831}
  (\bibinfo {year} {2001})}\BibitemShut {NoStop}%
\bibitem [{\citenamefont {Islas}\ and\ \citenamefont
  {Schober}(2005)}]{islas2005predicting}%
  \BibitemOpen
  \bibfield  {author} {\bibinfo {author} {\bibfnamefont {A.}~\bibnamefont
  {Islas}}\ and\ \bibinfo {author} {\bibfnamefont {C.}~\bibnamefont
  {Schober}},\ }\href@noop {} {\bibfield  {journal} {\bibinfo  {journal}
  {Physics of Fluids (1994-present)}\ }\textbf {\bibinfo {volume} {17}},\
  \bibinfo {pages} {031701} (\bibinfo {year} {2005})}\BibitemShut {NoStop}%
\bibitem [{\citenamefont {Toenger}\ \emph {et~al.}(2015)\citenamefont
  {Toenger}, \citenamefont {Godin}, \citenamefont {Billet}, \citenamefont
  {Dias}, \citenamefont {Erkintalo}, \citenamefont {Genty},\ and\ \citenamefont
  {Dudley}}]{toenger2015emergent}%
  \BibitemOpen
  \bibfield  {author} {\bibinfo {author} {\bibfnamefont {S.}~\bibnamefont
  {Toenger}}, \bibinfo {author} {\bibfnamefont {T.}~\bibnamefont {Godin}},
  \bibinfo {author} {\bibfnamefont {C.}~\bibnamefont {Billet}}, \bibinfo
  {author} {\bibfnamefont {F.}~\bibnamefont {Dias}}, \bibinfo {author}
  {\bibfnamefont {M.}~\bibnamefont {Erkintalo}}, \bibinfo {author}
  {\bibfnamefont {G.}~\bibnamefont {Genty}}, \ and\ \bibinfo {author}
  {\bibfnamefont {J.~M.}\ \bibnamefont {Dudley}},\ }\href@noop {} {\bibfield
  {journal} {\bibinfo  {journal} {Scientific reports}\ }\textbf {\bibinfo
  {volume} {5}} (\bibinfo {year} {2015})}\BibitemShut {NoStop}%
\bibitem [{\citenamefont {Onorato}\ \emph {et~al.}(2006)\citenamefont
  {Onorato}, \citenamefont {Osborne}, \citenamefont {Serio}, \citenamefont
  {Cavaleri}, \citenamefont {Brandini},\ and\ \citenamefont
  {Stansberg}}]{onorato2006extreme}%
  \BibitemOpen
  \bibfield  {author} {\bibinfo {author} {\bibfnamefont {M.}~\bibnamefont
  {Onorato}}, \bibinfo {author} {\bibfnamefont {A.}~\bibnamefont {Osborne}},
  \bibinfo {author} {\bibfnamefont {M.}~\bibnamefont {Serio}}, \bibinfo
  {author} {\bibfnamefont {L.}~\bibnamefont {Cavaleri}}, \bibinfo {author}
  {\bibfnamefont {C.}~\bibnamefont {Brandini}}, \ and\ \bibinfo {author}
  {\bibfnamefont {C.}~\bibnamefont {Stansberg}},\ }\href@noop {} {\bibfield
  {journal} {\bibinfo  {journal} {European Journal of Mechanics-B/Fluids}\
  }\textbf {\bibinfo {volume} {25}},\ \bibinfo {pages} {586} (\bibinfo {year}
  {2006})}\BibitemShut {NoStop}%
\bibitem [{\citenamefont {Dysthe}(1979)}]{dysthe1979note}%
  \BibitemOpen
  \bibfield  {author} {\bibinfo {author} {\bibfnamefont {K.~B.}\ \bibnamefont
  {Dysthe}},\ }in\ \href@noop {} {\emph {\bibinfo {booktitle} {Proc. Royal
  Society of London A: Mathematical, Physical and Engineering Sciences}}},\
  Vol.\ \bibinfo {volume} {369}\ (\bibinfo {organization} {The Royal Society},\
  \bibinfo {year} {1979})\ pp.\ \bibinfo {pages} {105--114}\BibitemShut
  {NoStop}%
\bibitem [{\citenamefont {Trulsen}\ and\ \citenamefont
  {Stansberg}(2001)}]{trulsen2001spatial}%
  \BibitemOpen
  \bibfield  {author} {\bibinfo {author} {\bibfnamefont {K.}~\bibnamefont
  {Trulsen}}\ and\ \bibinfo {author} {\bibfnamefont {C.~T.}\ \bibnamefont
  {Stansberg}},\ }in\ \href@noop {} {\emph {\bibinfo {booktitle} {The Eleventh
  International Offshore and Polar Engineering Conference}}}\ (\bibinfo
  {organization} {International Society of Offshore and Polar Engineers},\
  \bibinfo {year} {2001})\BibitemShut {NoStop}%
\bibitem [{\citenamefont {Trulsen}\ and\ \citenamefont
  {Dysthe}(1996)}]{trulsen1996modified}%
  \BibitemOpen
  \bibfield  {author} {\bibinfo {author} {\bibfnamefont {K.}~\bibnamefont
  {Trulsen}}\ and\ \bibinfo {author} {\bibfnamefont {K.~B.}\ \bibnamefont
  {Dysthe}},\ }\href@noop {} {\bibfield  {journal} {\bibinfo  {journal} {Wave
  Motion}\ }\textbf {\bibinfo {volume} {24}},\ \bibinfo {pages} {281} (\bibinfo
  {year} {1996})}\BibitemShut {NoStop}%
\bibitem [{\citenamefont {Chabchoub}\ \emph
  {et~al.}(2013{\natexlab{b}})\citenamefont {Chabchoub}, \citenamefont
  {Hoffmann}, \citenamefont {Onorato}, \citenamefont {Genty}, \citenamefont
  {Dudley},\ and\ \citenamefont {Akhmediev}}]{chabchoub2013hydrodynamic}%
  \BibitemOpen
  \bibfield  {author} {\bibinfo {author} {\bibfnamefont {A.}~\bibnamefont
  {Chabchoub}}, \bibinfo {author} {\bibfnamefont {N.}~\bibnamefont {Hoffmann}},
  \bibinfo {author} {\bibfnamefont {M.}~\bibnamefont {Onorato}}, \bibinfo
  {author} {\bibfnamefont {G.}~\bibnamefont {Genty}}, \bibinfo {author}
  {\bibfnamefont {J.~M.}\ \bibnamefont {Dudley}}, \ and\ \bibinfo {author}
  {\bibfnamefont {N.}~\bibnamefont {Akhmediev}},\ }\href@noop {} {\bibfield
  {journal} {\bibinfo  {journal} {Phys. Rev. Lett.}\ }\textbf {\bibinfo
  {volume} {111}},\ \bibinfo {pages} {054104} (\bibinfo {year}
  {2013}{\natexlab{b}})}\BibitemShut {NoStop}%
\bibitem [{\citenamefont {Trulsen}(2007)}]{trulsen2007weakly}%
  \BibitemOpen
  \bibfield  {author} {\bibinfo {author} {\bibfnamefont {K.}~\bibnamefont
  {Trulsen}},\ }in\ \href@noop {} {\emph {\bibinfo {booktitle} {Geometric
  Modelling, Numerical Simulation, and Optimization}}}\ (\bibinfo  {publisher}
  {Springer},\ \bibinfo {year} {2007})\ pp.\ \bibinfo {pages}
  {191--209}\BibitemShut {NoStop}%
\bibitem [{\citenamefont {Ducrozet}\ \emph {et~al.}(2012)\citenamefont
  {Ducrozet}, \citenamefont {Bonnefoy}, \citenamefont {Le~Touz{\'e}},\ and\
  \citenamefont {Ferrant}}]{ducrozet2012modified}%
  \BibitemOpen
  \bibfield  {author} {\bibinfo {author} {\bibfnamefont {G.}~\bibnamefont
  {Ducrozet}}, \bibinfo {author} {\bibfnamefont {F.}~\bibnamefont {Bonnefoy}},
  \bibinfo {author} {\bibfnamefont {D.}~\bibnamefont {Le~Touz{\'e}}}, \ and\
  \bibinfo {author} {\bibfnamefont {P.}~\bibnamefont {Ferrant}},\ }\href@noop
  {} {\bibfield  {journal} {\bibinfo  {journal} {European Journal of
  Mechanics-B/Fluids}\ }\textbf {\bibinfo {volume} {34}},\ \bibinfo {pages}
  {19} (\bibinfo {year} {2012})}\BibitemShut {NoStop}%
\bibitem [{\citenamefont {Slunyaev}\ and\ \citenamefont
  {Shrira}(2013)}]{slunyaev2013highest}%
  \BibitemOpen
  \bibfield  {author} {\bibinfo {author} {\bibfnamefont {A.~V.}\ \bibnamefont
  {Slunyaev}}\ and\ \bibinfo {author} {\bibfnamefont {V.~I.}\ \bibnamefont
  {Shrira}},\ }\href@noop {} {\bibfield  {journal} {\bibinfo  {journal} {J.
  Fluid Mechanics}\ }\textbf {\bibinfo {volume} {735}},\ \bibinfo {pages} {203}
  (\bibinfo {year} {2013})}\BibitemShut {NoStop}%
\bibitem [{\citenamefont {Akhmediev}\ \emph {et~al.}(2011)\citenamefont
  {Akhmediev}, \citenamefont {Ankiewicz}, \citenamefont {Soto-Crespo},\ and\
  \citenamefont {Dudley}}]{akhmediev2011roguespectra}%
  \BibitemOpen
  \bibfield  {author} {\bibinfo {author} {\bibfnamefont {N.}~\bibnamefont
  {Akhmediev}}, \bibinfo {author} {\bibfnamefont {A.}~\bibnamefont
  {Ankiewicz}}, \bibinfo {author} {\bibfnamefont {J.}~\bibnamefont
  {Soto-Crespo}}, \ and\ \bibinfo {author} {\bibfnamefont {J.~M.}\ \bibnamefont
  {Dudley}},\ }\href@noop {} {\bibfield  {journal} {\bibinfo  {journal} {Phys.
  Lett. A}\ }\textbf {\bibinfo {volume} {375}},\ \bibinfo {pages} {541}
  (\bibinfo {year} {2011})}\BibitemShut {NoStop}%
\bibitem [{\citenamefont {Randoux}\ \emph {et~al.}(2015)\citenamefont
  {Randoux}, \citenamefont {Suret},\ and\ \citenamefont
  {El}}]{randoux2015identification}%
  \BibitemOpen
  \bibfield  {author} {\bibinfo {author} {\bibfnamefont {S.}~\bibnamefont
  {Randoux}}, \bibinfo {author} {\bibfnamefont {P.}~\bibnamefont {Suret}}, \
  and\ \bibinfo {author} {\bibfnamefont {G.}~\bibnamefont {El}},\ }\href@noop
  {} {\bibfield  {journal} {\bibinfo  {journal} {arXiv preprint
  arXiv:1512.04707}\ } (\bibinfo {year} {2015})}\BibitemShut {NoStop}%
\end{thebibliography}%

\end{document}